\documentstyle[prd,aps,12pt]{revtex}

\tighten

\newcommand{\be}{\begin{eqnarray}}
\newcommand{\ee}{\end{eqnarray}}
\newcommand{\ben}{\begin{eqnarray*}}
\newcommand{\een}{\end{eqnarray*}}
\newcommand{\nn}{\nonumber \\ }

\newcommand{\da}{\dagger}
\newcommand{\Cdot}{\! \cdot \!}
\newcommand{\Not}{\! \not \!}

\newcommand{\pr}{\partial}

\newcommand{\barN}{\overline{N}}

\newcommand{\baru}{\overline{u}}

\begin{document}

\draft
\title{The reaction $\pi N \rightarrow \pi\pi N$ above
threshold in heavy-baryon chiral perturbation theory}

\author{J. Zhang and 
\underline{N. Mobed}\footnote{ \noindent \underline{Corresponding Author} \\ 
E-mail: mobed@meena.cc.uregina.ca\\ Fax: (306)-585-4894 \\
Tel: (306)-585-4359}\\
Department of Physics\\ 
University of Regina\\
Regina, SK S4S 0A2, Canada \\[1em]
M. Benmerrouche\\ 
Saskatchewan Accelerator Laboratory \\
University of Saskatchewan\\
Saskatoon, SK S7N 5C6, Canada}

\maketitle

\begin{abstract}

We study the reactions $\pi^{\pm} p \rightarrow \pi^{\pm} \pi^{+} n$ and 
$\pi^{-} p \rightarrow \pi^{o} \pi^{o} n$ in heavy-baryon chiral perturbation
theory of chiral order three from threshold up to
pion laboratory kinetic energies of $400$ MeV. We find that
the contributions from
amplitudes of chiral order three are large and play an essential
role in reproducing the experimental data. 
 
\end{abstract}


\newpage

The reaction $N(\pi,2\pi)N$ has been the subject of a 
number of experimental\cite{manl84,poca93} and 
theoretical\cite{oset85,jake92,jens97,beri92,bkm95,bkm97} 
investigations. The reaction is of interest in connection with various 
aspects of chiral symmetry and its spontaneous breaking, including a 
determination of $\pi$-$\pi$ scattering amplitude and the study of 
non-linear realization of chiral symmetry. The theoretical work on 
the reaction $N(\pi,2\pi)N$ may be broadly divided into 
two categories. In one category, calculations are 
based on a tree-level chiral Lagrangian and include meson and baryon 
resonances as explicit degrees of freedom \cite{oset85,jake92,jens97}. 
While these models capture the physical essence of the 
reaction under consideration, they are not consistent with the 
power-counting scheme of Chiral Perturbation Theory (ChPT) 
\cite{gass88,krau90}. 
The second category includes calculations which are fully within the 
framework of the baryon ChPT. In \cite{beri92} the lowest order 
($O(q)$) relativistic 
Lagrangian was employed, while in \cite{bkm97} the calculations were 
performed in the next-to-leading order ($O(q^2)$) relativistic 
framework. The power counting ambiguities of the relativistic formalism
are discussed in \cite{gass88}. Heavy-Baryon Chiral Perturbation Theory
(HBChPT) circumvents these difficulties \cite{bern95a,ecke95}. 
The most comprehensive 
study of the reaction $N(\pi,2\pi)N$ at the 
threshold was carried out in 
\cite{bkm95}, where a complete $O(q^3)$ calculation was 
performed in HBChPT. The purpose of 
this paper is to investigate the reaction 
$N(\pi,2\pi)N$ beyond the threshold in HBChPT up to $O(q^3)$.
Technically, the threshold and beyond the threshold 
calculations differ in two respects.  First, as one moves away from 
threshold the number of Feynman diagrams  
increases significantly.  Second, the evaluation of loop integrals 
becomes more involved.  

Our starting point is the low energy expansion of the $\pi N$ 
chiral Lagrangian in HBChPT
as described in detail in Ref.\cite{ecke96a}. The effective chiral 
Lagrangian may be expanded as
\be
{\cal L}_{\rm eff}=
{\cal L}_{\pi}^{(2)} + {\cal L}_{\pi}^{(4)} + \cdots
+{\cal L}^{(1)}_{\pi N} + {\cal L}^{(2)}_{\pi N} +
{\cal L}^{(3)}_{\pi N} +\cdots,
\ee
where the superscripts denote the chiral order in the standard
power counting scheme\cite{bern95a,ecke95}. In order
to obtain amplitudes of $O(q^3)$, the pion and the pion-nucleon
Lagrangians may be truncated at $O(q^4)$ and $O (q^3)$,
respectively. The pion 
${\cal L}_{\pi}^{(i)}$, and the pion-nucleon, ${\cal L}^{(i)}_{\pi N}$ parts
of the Lagrangian are discussed in detail in \cite{gass88} and 
\cite{ecke96a}, 
respectively. In the absence of external fields and in the isospin symmetric 
limit, the pertinent pieces of the Lagrangian assume the following form:
\be
{\cal L}_\pi ^{(2)} & = & \frac{F^2}{4} \langle u \cdot u \rangle + 
\frac{1}{2} m_\pi^2 F^2 \langle U \rangle  \, , 
\\ \rule{0mm}{8mm}
{\cal L}_\pi ^{(4)} & = & \frac{1}{16} \left \{
4 l_1 \langle u \cdot u \rangle ^2 + 4 l_2  
\langle  u_\mu u^\nu \rangle \langle  u^\mu u_\nu \rangle +
l_3 \langle  \chi_+ \rangle ^2  
\right.
\nn \rule{0mm}{8mm} & & \hspace{6mm}
\left.
+ l_4 ( 2 \langle  \chi_+ \rangle \langle  u \cdot u \rangle  + 2 \langle \chi_-^2\rangle 
- \langle  \chi_- \rangle ^2 )\, \right \} \, ,
\\ \rule{0mm}{8mm}
{\cal L}_{\pi N}^{(1)} & = & 
\barN_v \left ( \vphantom{\frac{1}{2}} i v \Cdot \nabla + g_A S \Cdot u \right ) N_v \, ,
\\ \rule{0mm}{8mm}
{\cal L}_{\pi N}^{(2)} & = & \barN_v \, \left \{
- \frac{1}{2M} 
\left ( \nabla^2 + i g_A \{ u \Cdot v , S \Cdot \nabla \} \right ) 
+ \frac{a_1}{M} \langle u \Cdot u \rangle 
+ \frac{a_2}{M} \langle (u \Cdot v)^2 \rangle
\right.
\nn \rule{0mm}{8mm} & & \hspace{8mm}
+ \frac{m_\pi^2 a_3}{M} \langle U + U^\da \, \rangle
+\frac{m_\pi^2 a_4}{M} \left ( U + U^\da - \frac{1}{2} \langle U + U^\da \,\rangle \right ) 
\nn \rule{0mm}{8mm} & & \hspace{8mm}
\left.
+ \frac{i a_5}{M} \epsilon^{\mu \nu \alpha \beta} 
v_\alpha S_\beta \, u_\mu u_\nu  \right \} N_v \, ,
\\ \rule{0mm}{8mm}
{\cal L}^{(3)}_{\pi N} & = & \frac{g_A} {8 M^2} \, \barN_v \, 
\left [ \vphantom{\frac{1}{2}} \nabla_\mu, 
\left [ \vphantom{\barN_v} \nabla^\mu, S \Cdot u \right ] \right ] \, N_v
\nn \rule{0mm}{8mm} & & 
+ \frac{1}{2M^2} \barN_v \, 
\left \{\vphantom{\frac{g_A^2}{8}} - 
\left( a_5 - \frac{1 - 3 g_A^2} {8} \right)
 \, u_\mu u_\nu
\epsilon^{\mu \nu \alpha \beta} S_\beta \nabla_\alpha
+ \frac{g_A}{2} S \Cdot \nabla u \Cdot \nabla 
\right.
\nn \rule{0mm}{8mm} & & \hspace{18mm} \, 
\left. 
- \frac{g_A^2}{8} \left \{v \Cdot u, u_\mu \right \} 
\epsilon^{\mu \nu \alpha \beta} v_\alpha S_\beta \nabla_\nu + \mbox{h.c.} 
\right \} 
N_v + \frac{1}{(4 \pi F)^2} \barN_v \left ( \sum_{i=1}^{24} b_i O_i \right ) 
N_v.
\ee
Here, $S^{\mu}=\frac{i} {2} \gamma _5 \sigma^{\mu \nu} v_{\nu}$ is the
spin matrix, 
$U=u^2=\exp (i \vec{\tau} \Cdot \vec{\pi} / F)$ is the SU(2) pion field with
$F$ being the pion decay constant, and $N_v(x)$ is the 
velocity-dependent nucleon field defined in terms of the Dirac spinor, 
$\psi(x)$, as
\be
N_v(x)=e^{i M v \cdot x} \frac{1}{2} (1 + \Not v) \psi(x), 
\ee
where $M$ is the nucleon mass, and the four-velocity $v^{\mu}$ satisfies the 
constraint $v^2 =1$. 
The last term in ${\cal L}^{(3)}_{\pi N}$ involving the Low Energy 
Constants (LECs) $b_i$ 
is usually referred to as the counterterm Lagrangian. Some of the 
coefficients $l_i$ and $b_i$ are divergent in order 
to absorb the divergences of the one-loop diagrams which start contributing 
at order $O(q^3)$. In the dimensional regularization scheme, 
these low energy constants $l_i$ and $b_i$ can be decomposed 
as\cite{dono92}: 
\be
l_i = l_i^r (\mu) + \gamma_i \Lambda,\quad 
b_i = b_i^r (\mu) + (4 \pi)^2 \beta_i \Lambda,
\ee
and 
\be
\Lambda = {1 \over 32\pi^2}\left [ {2\over (d-4)} - \ln (4\pi) 
-\Gamma'(1) - 1\right ],
\label{bi}
\ee
where $l_i^{r}(\mu)$ and $b_i^r(\mu)$ are the renormalized values of 
$l_i$ and $b_i$ at 
the scale $\mu$.
Ecker and Moj\v{z}i\v{s}\cite{ecke96a} have worked out the counterterms and 
their $\beta$--functions involved in ${\cal L}^{(3)}_{\pi N}$ 
and we refer to their paper for the 
explicit form of the operators $O_i$ and the coefficients $\beta_i$.
The coefficients $\gamma_i$ are given in \cite{gass88,mojz97} .
We have also used the standard notation for 
the operators $\chi = (m_u + m_d) B_0 = m_\pi^2$, 
$u_\mu = i (u^\da \pr_\mu u - u \pr_\mu u^\da )$, $\nabla = \pr + 
\Gamma$ and $\Gamma_\mu = \frac{1}{2} ( u^\da \pr_\mu u + u \pr_\mu u ^\da )$. 
By expanding the field $U=u^2$ in powers of the pion field and substituting the
result in the effective Lagrangian, we obtain an explicit form of the
Lagrangian suitable for deriving the Feynman rules.
To a given chiral order, the transition amplitude ${\cal T}_{fi}$ receives 
contributions from all the possible connected Feynman diagrams up to 
that order.  The (non-vanishing) topologically distinct diagrams of up to 
$O(q^3)$ contributing to the reaction $N(\pi, 2 \pi)N$ are given in Fig.~1.
The self-energy diagrams are not shown.  
It is worth emphasizing that the actual number of 
diagrams for any particular physical channel is much greater than 
that of the 
ones displayed in Fig.~1. The individual Feynman graphs of $O(q^3)$ 
are not necessarily finite, but the total transition amplitude is 
finite at each order.
 
Before presenting our results we have to make a few remarks  
on the input parameters of our calculations. 
For the reaction under consideration up to $O(q^3)$ the 
following vertices and LECs are needed: ${\cal L}^{(2)}_{\pi}$ 
involving $m_{\pi}$ and $F_{\pi}$, ${\cal L}^{(4)}_{\pi}$ with LECs 
$l_i$ $(i=1,\cdots,4)$, 
${\cal L}^{(1)}_{\pi N}$ containing $g_A$, ${\cal L}^{(2)}_{\pi N}$ 
containing $M$ and $a_1$ through $a_5$, 
and finally ${\cal L}^{(3)}_{\pi N}$, containing 24 LECs $b_i$ of which 14 
can, in principle, contribute to the reaction under consideration.  
Of the 14 LECs, 5 combinations containing 9 LECs contribute to the 
$\pi N$ elastic scattering; these combinations have been recently determined
\cite{mojz97}. The remaining LECs needed in our calculation 
are $b_4, b_{11}, b_{12}, b_{13}, b_{14}, b_{17}$ .  
In this work, we treat the finite part of the 
unknown LECs as free parameters and denote them by $\tilde{b}_i$.
The numerical values of $a_i$ and $\tilde{b_i}$ are listed in
Tables \ref{table1} and \ref{table2}, respectively.  
Throughout the calculations the physical values of the quantities 
$m_{\pi}$, $F_{\pi}$, $g_A$ and $M$ are used.  
As discussed in \cite{mojz97} this amounts to corrections 
of $O(q^3)$ and higher. The corrections of $O(q^3)$ are explicitly 
included in the amplitudes of the same order. 

In HBChPT the transition amplitude for the reaction
\be
\pi(q) + N(p=Mv+l_i) \longrightarrow \pi(q_1) + \pi(q_2) + N (p'=Mv + l_f)
\ee 
can be written as
\be
{\cal T}_{fi} = 
\bar{u}_v^{(\alpha_f)}(l_f) \, {\cal A} \, u_v^{(\alpha_i)}(l_i) ,
\label{amp}
\ee
in which $u_v^{(\alpha)}(l)$ denotes the heavy baryon spinor with 
residual momentum $l$ and spin projection $\alpha$. We find that
the amplitude ${\cal A}$ has the following Lorentz structure 
\be
{\cal A} = 
A_0\,S \cdot q + A_1\,S\cdot q_1 + A_2\,S\cdot q_2 + i 
\epsilon_{\alpha \beta \mu \nu} v^{\alpha} q^\beta q_1^\mu q_2^\nu B,
\label{tfi}
\ee
where $A_i$ $(i = 0, 1, 2)$ and $B$ are invariant functions of external momenta.
 
In order to make contact with the experimental data, 
we need to relate the frame-dependent HBChPT amplitudes to the 
relativistic invariant amplitudes. 
The relativistic amplitude for the reaction $N(\pi,2\pi)N$ can be parameterized
as \cite{jens97,beri92}:
\be
{\cal M}_{fi} = \baru(p') \, \gamma_5 
\left [ \vphantom{A^2}
f_1 + f_2 \, \Not q_1 + f_3 \, \Not q_2 + f_4 \, \Not q_1 \, 
\Not q_2 \right] u(p).
\label{Re_Am}
\ee
The use of (\ref{tfi}) and (\ref{Re_Am}), along with the matching 
condition introduced in \cite{ecke97} results in the following relations between
the relativistic and heavy-baryon amplitudes: 
\be
f_1 & = & M \left (1 - \frac{t}{4 M^2}  \right ) 
\left (A_0 - 2 B \, q_1 \Cdot q_2  \right )
+ \frac{1}{2} (A_0 + A_1) \, v \Cdot q_1 + 
\frac{1}{2} (A_0 + A_2) \, v \Cdot q_2 
\nn  \rule{0mm}{8mm}& & 
+ \left [ v \Cdot q_1 \, q \Cdot q_2 -  v \Cdot q_2 \, q \Cdot q_1 + 
v \Cdot (q_2 - q_1) \, ( m_\pi^2 + q_1 \Cdot q_2 )  \right ] B ,
\label{f_1}
\\ \rule{0mm}{8mm}
f_2 & = & - \frac{1}{2} (A_0 + A_1) \,
- \left [2 M \, v \Cdot q_2 + q_2 \Cdot ( q - q_1 ) - m_\pi^2  \right ]  B , 
\\ \rule{0mm}{8mm}
f_3 & = & - \frac{1}{2} (A_0 + A_2) \,
+ \left [ 2 M \, v \Cdot q_1 + q_1 \Cdot ( q  - q_2 ) - m_\pi^2  \right ]  B , 
\\ \rule{0mm}{8mm}
f_4 & = & 2 M \left (1 - \frac{t}{4 M^2}  \right ) B,
\label{f_4}
\ee
where $t= (p - p')^2$.
At threshold the relativistic transition amplitude takes a simple 
form:
\be
{\cal M}_{fi} = a \, {\chi^{(\alpha_f)}}^\da \,
\vec \sigma \Cdot \vec q \, \chi^{(\alpha_i)} ,
\ee
where
\be
a = \frac{1}{2 M} 
\left [ m_\pi \, (f_2  + f_3 ) - f_1 - m_\pi^2 \, f_4  \right ]  .
\label{a}
\ee
The threshold amplitudes corresponding to a well-defined isospin 
channel, commonly referred to as $a^{3,2}$ and 
$a^{1,0}$, 
can be decomposed in terms of physical amplitudes \cite{jens97}: 
\be 
a^{1,0} & = &\frac{{3}}{\sqrt{2}} \, a^{\pi^{0}\pi^{0}n} 
- \frac{1}{\sqrt{2}} \, a^{\pi^{+}\pi^{+}n},\\ 
a^{3,2} & = & \frac{\sqrt{5}}{2} \,a^{\pi^{+}\pi^{+}n},
\ee
where $a^{\pi^{0}\pi^{0}n}$ and $a^{\pi^{+}\pi^{+}n}$ are the 
threshold amplitudes for the physical reactions  
$\pi^- p \rightarrow \pi^0 \pi^0 n$ and $\pi^+ p \rightarrow \pi^+ 
\pi^+n$ respectively. Furthermore, the 
amplitudes $D_{1},D_{2}$ defined by Bernard, Kaiser, 
Mei{\ss}ner (BKM)\cite{bkm95} are related to 
$a^{i,j}$ by
\be
D_1 & = & \frac{1}{\sqrt{10}} \, a^{3,2},
\\ \rule{0mm}{8mm}
D_2 & = & - \frac{2}{3} \, \frac{a^{3,2}}{\sqrt{10}} - \frac{1}{3 } \, a^{1,0}.
\ee

Our numerical results for $D_1$ and $D_2$ up to chiral order $O(q^3)$
are listed in Tables \ref{table3} and \ref{table4}. 
Clearly the contribution of
non-leading terms to the threshold amplitudes are substantial, and the results 
seem to be sensitive to the values of the unknown LECs. 
Our calculated values of $D_1$ and $D_2$ seem to be in 
fair agreement with the data. These statements are somewhat
tentative in that the use of the actual values of the unknown LECs may
lead to a different set of conclusions. The HBChPT calculations of
BKM, on the other hand, are 
in impressive agreement with the data. In principle, the difference
between our predictions and those of BKM is a consequence of using
different values of LECs $\tilde b_i$. In their work, BKM estimated
the contribution of the terms containing the unknown LECs using the 
resonance saturation method\cite{bkm95}. As such, there are no free 
parameters in their calculations. 
It is conceivable that with a systematic variation of the
unknown LECs we may also obtain a good agreement with the data. Finally,
we note that the calculations of Ref.\cite{jens97}, while not fully 
consistent with the principles of chiral perturbation theory,
yield results which are in close agreement with our calculations.

We now turn our attention to the total cross sections.
The spin-averaged cross-section is given by 
\be
\sigma = \frac{2M} { 4 s \, [(p \cdot q )^2 - M^2 m_\pi^2 ]^{1/2} } 
\int  \overline{|{\cal T}_{fi}|^2}
\frac{d^3 p'}{(2 \pi)^3 } \frac{M}{E_{p'}} 
\frac{d^3 q_1}{(2 \pi)^3 2 \omega_{q_1}} 
\frac{d^3 q_2}{(2 \pi)^3 2 \omega_{q_2}} 
(2 \pi)^4 \delta^4 (P_f - P_i),
\label{xsec}
\ee
where $s$ is a statistical factor accounting for the
existence of identical particles in the final state. 
Using (\ref{tfi}), we obtain  
\be
\overline{|{\cal T}_{fi}|^2} = C(p) C(p') 
\left \{ \frac{1}{4} (v \Cdot A \, v \Cdot A^* - |A|^2 )
+ \left (\epsilon_{\alpha \beta \mu \nu} v^{\alpha} q^\beta 
q_1^\mu q_2^\nu \right)^2 |B|^2 \right \},
\label{tfi2}
\ee
where $C(p) \equiv \displaystyle{\frac{v \Cdot p + M}{2 M} }$. To compute the 
cross-section, one has to integrate (\ref{tfi2}) over an 
appropriate phase-space.  The phase-space integration was carried out 
numerically using a Monte-Carlo routine GENBOD from the CERN program 
library\cite{cern}. It is worth noting that while the transition 
amplitude is evaluated in the isospin limit, physical masses are used for 
the reaction kinematics.

The calculations have been performed for the Lab incident pion kinetic energy,
$T_\pi$, from threshold up to $400$ MeV.  From a theoretical
point of view, chiral perturbation theory is expected to be more reliable
at low energies. Hence, for each reaction, we shall present our results for 
two kinematic domains: from threshold to $200$ MeV, and 
from threshold to $400$ MeV.
Furthermore, to facilitate comparison among different chiral orders,
results will be displayed order by order in the chiral expansion.
Finally, in order to study  the sensitivity of the 
results to the values of the unknown LECs, 
$\tilde b_i$, we have performed three sets of calculation for each reaction:
all unknown $\tilde b_i = 0$, unknown $\tilde b_i = 10$, 
and unknown $\tilde b_i = -10$. 
The choice of the range of variation of the unknown $\tilde b_i$ is 
based on the reported values of the known LECs $\tilde b_i$ as listed in
Table \ref{table2}. Some of the known LECs in Table \ref{table2} have rather
large uncertainties. It was found that the cross-sections are 
not noticeably affected by these uncertainties. 
Our results are illustrated in Figs.~2 and 3. 

Our main results can be summarized as follows. 
The leading order amplitudes consistently underestimate the experimental
results by a large amount over most of the kinematic range considered 
in our calculations. The inclusion of terms of $O(q^2)$ produces only
a small change in the cross-section. The contribution of terms of 
$O(q^3)$, on the other hand, is quite large and in general results
in an improved agreement with the data.
From a phenomenological point of view these observations can 
be regarded as a success for HBChPT. However, from a theoretical 
point of view, the fact that the contribution of terms of order 
$O(q^3)$ is much larger than that of the lower order terms, 
is indicative of the slow rate of convergence of HBChPT. The slow 
convergence of HBChPT has also been observed \cite{bern95a,mojz97} for
processes other than $N(\pi, 2\pi)N$. 
The lack of knowledge about the values of a number of LECs,
$\tilde b_i$, is the main source of uncertainty in our calculations.
Our results suggest that variations in $\tilde b_i$ can produce 
quite noticeable effects in observables considered here. 
Clearly a systematic investigation is necessary in order to determine
or constrain the unknown LECs. So far as the reaction $N(\pi, 2\pi)N$
is concerned, one could try to fit the total cross-section in 
conjunction with a number of more sensitive differential cross-section
and angular correlation data. 
These investigations are beyond the scope of the present work and
will be the subject of a future publication. 

To conclude, we note that our quantitative results for the total
cross-section are in fair agreement
with previous calculations for the reaction $N(\pi,2\pi)N)$
\cite{oset85,jake92,jens97,beri92,bkm97}.
The fundamental difference between the present work and the previous
ones is the calculational framework employed. Our calculations are
performed in HBChPT and hence obey a consistent chiral power
counting scheme.

This work has been supported by the Natural Sciences and Engineering Research 
Council of Canada.

\include{refs}

\pagebreak


\newpage
Fig.~1: {Topologically distinct graphs contributing to the $\pi N 
\rightarrow \pi\pi N$ reaction up to chiral order $O(q^3)$. 
Self-energy diagrams 
and diagrams which give zero contribution are not shown.} 

Fig.~2: {The total cross section in the threshold region for 
$\pi^{+} p \rightarrow \pi^{+} \pi^{+} n$, 
$\pi^{-} p \rightarrow \pi^{+} \pi^{-} n$ and 
$\pi^{-} p \rightarrow \pi^{o} \pi^{o} n$ reactions. 
The long dashed curve is our prediction to chiral order $O(q)$, 
the dashed-dotted curve corresponds to $O(q) + O(q^2)$, the solid curve is the 
complete calculation  $O(q) + O(q^2) + O(q^3)$ 
with all unknown LECs set to zero. The dotted and short 
dashed curves show the 
sensitivities to upward and downward shifts of all the unknown LECs 
by $10$ units, respectively.
The experimental data are represented by the symbols: Filled 
diamonds: \protect\cite{sevi91}, 
empty triangles: \protect{\cite{kern90}, Filled squares: \protect\cite{kirz62}, 
Filled circles: \protect\cite{kern89}, Empty circles: \protect\cite{bjor80}, 
Stars: \protect\cite{sevi97},
Filled Triangles: \protect\cite{lowe91}.}

Fig.~3: {Same as Fig.~2 but for a wider kinematic range.}

\newpage


\begin{table}

\caption{The LECs contributing to amplitudes of 
$O(q^2)$ or higher \protect\cite{mojz97} }
\label{table1}
\begin{center}
\begin{tabular}{|c|c|c|c|}
$~~~~~~~a_1~~~~~~$  &   $a_2~~~~~~~~$  &   $a_3~~~~~~~$  &   $a_5~~~~~~~~$ \\
\hline
$~~~~~~~-2.60 \pm 0.03~~~~~~~$  & $1.40 \pm 0.05 ~~~~~~~~$ & $-1.00 \pm 0.06~~~~~~~$
& $3.30 \pm 0.05 ~~~~~~~~$  \\
\end{tabular}
\end{center}

\caption{The LECs contributing to amplitudes of $O(q^3)$ 
\protect\cite{gass88,mojz97}}
\label{table2}
\begin{center}
\begin{tabular}{|c|c|c|c|}
 $~~~~~~~~~~~~~~\bar l_1~~~~~~~~~~$    &   $\bar l_2~~~~~~~~~$   &    
$\bar l_3~~~~~~~~~~$  &   $\bar l_4~~~~~~~~~$ \\
\hline
$ ~~~~~~~~~-2.3 \pm 3.7~~~~~~~$ & $ 6.0 \pm 1.3 ~~~~~~~~~$ & $ 2.9 \pm 2.4 ~~~~~~~~~$ & 
$ 4.3 \pm 0.9 ~~~~~~~~~$ 
\\
\end{tabular}
\end{center}

\begin{center}
\begin{tabular}{|c|c|c|c|c|}
 $~~~~~~~~\tilde b_1+\tilde b_2~~~~~~~~$ &  $\tilde b_3~~~~~$  & $ \tilde b_6~~~~~ $
 &  $\tilde b_{15}-\tilde b_{16} ~~~~~$  
 & $ \tilde b_{19} ~~~~~$  \\ \hline
$ 2.4 \pm 0.3$ & $ -2.8 \pm 0.6~~~~~ $ & $ 1.4 \pm 0.3~~~~~ $
& $ -6.1  \pm 0.4 ~~~~~~~~$ & $ -2.4 \pm 0.2 ~~~~~~ $ \\
\end{tabular}
\end{center}
\end{table}
\begin{table}[ht]
\caption{Calculated values of $D_1$ and $D_2$ in this work}
\label{table3}
\begin{center}
\begin{tabular}{|c|c|c|c|c|c|}
& $O(q)~~~$ & $O(q^2)~~~~$ & $O(q^3)$, ($\tilde b_i$ = 0) ~~~& 
$O(q^3)$, ($\tilde b_i = 10$) ~~~& 
$O(q^3)$, ($\tilde b_i = -10$) ~~~\\ \hline
$D_1$(fm$^3$) & 2.50 ~~& 2.14~~ & 2.04 ~~& 0.50 & 4.46 \\ \hline
$D_2$(fm$^3$) & $- 7.70~~~~$  & $- 6.96 ~~~~$ & $- 10.88~~~~~$ & $- 8.92~~~$ & $-13.95~~~~$  \\ 
\end{tabular}
\end{center}
\end{table}
\begin{table}[ht]
\caption{Values of $D_1$ and $D_2$}
\label{table4}
\begin{center}
\begin{tabular}{|c|c|c|c|c|}
& Experiment \cite{burk91}~~~~~~& Present Work~~~~~~&
Theory \cite{bkm95}~~~~~~& Theory \cite{jens97}~~~~~~\\ 
&  & $O(q^3)$, ($\tilde b_i = 0$) ~~~~&  & \\
\hline
$D_1(\mbox{fm}^3)$ & $2.26 \pm 0.12~~$ & 2.04 ~~~~& $2.65 \pm 0.24 ~~~~$ & 1.87 ~~~~\\
\hline
$D_2(\mbox{fm}^3)$ & $- 9.05 \pm 0.36 ~~~~$ & $- 10.88~~~~~~~$ & $- 9.06 \pm 1.05 ~~~~~~~$ & 
$- 10.58 $ ~~~~~~~\\
\end{tabular}
\end{center}
\end{table}

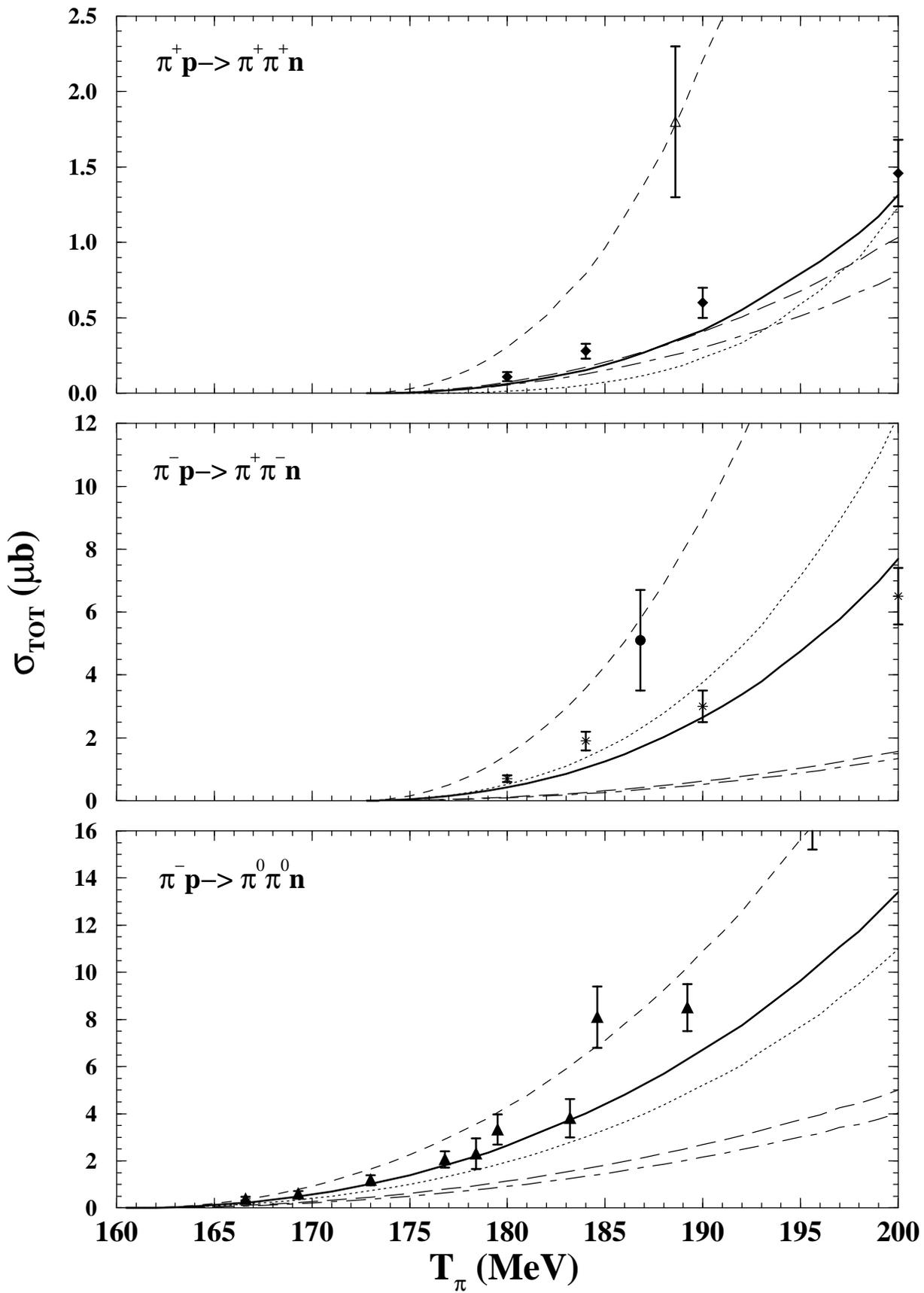
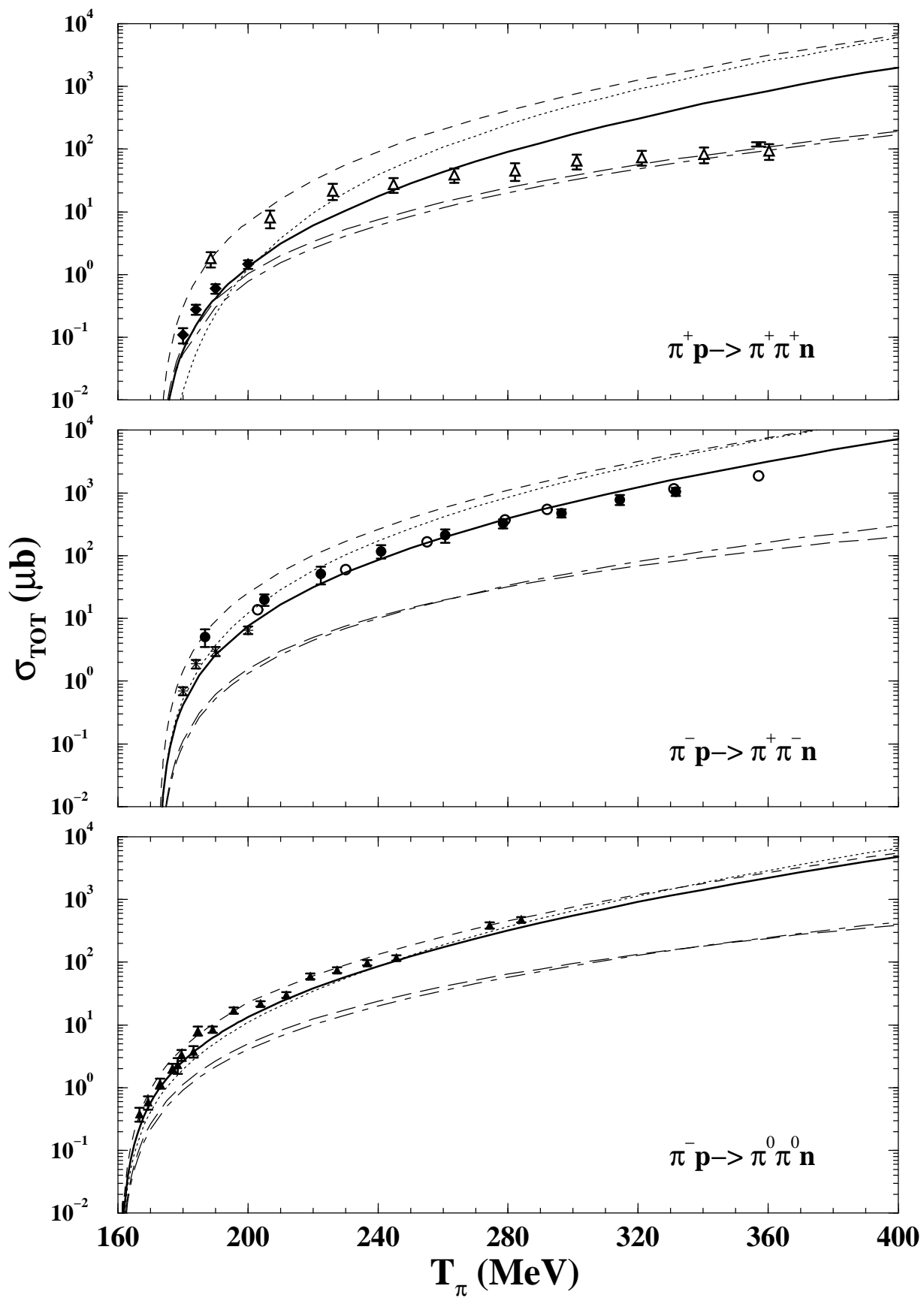
\end{document}